\documentclass[12pt]{article}%
\usepackage{citesort}
\usepackage{amsmath}
\usepackage{graphicx}%
\usepackage{amsfonts}%
\usepackage{amssymb}
\begin{document}

\title{Charged pions multiplicities at the NA49 energy}
\author{A. Bzdak\thanks{e-mail: Adam.Bzdak@ifj.edu.pl}\\Institute of Nuclear Physics, Polish Academy of Sciences\\Radzikowskiego 152, 31-342 Krakow, Poland}
\maketitle

\begin{abstract}
The wounded quark-diquark model predictions for charged pions multiplicities
in $PbPb$ and $pPb$ collisions in the central rapidity region at $\sqrt
{s}=17.3$ GeV c.m. energy are presented.

\vskip 1cm
%\noindent PACS: 25.75.-q, 25.75.Ag, 21.65.Qr\newline Keywords: NA49, heavy ion
%collisions, diquark, wounded nucleons

\end{abstract}

\textbf{1.} The NA49 collaboration published precise results\footnote{Total
systematic uncertainty of $2.0\%$ (quadratic sum) and $4.8\%$ (upper limit).}
on inclusive production of charged pions in $pp$ collisions at $\sqrt{s}=17.3$
GeV \cite{na49}.

This measurement allows to investigate the consequences of the wounded
quark-diquark model \cite{bb-AA} for particles production in the central
rapidity region of $pPb$ and $PbPb$ collisions at the same energy. We conclude
that the model provides rather precise predictions (at the level of $2-3\%$)
for the production of charged pions in nuclear collisions.

\textbf{2.} We follow closely the Ref. \cite{lhc} where the predictions of the
model in $pPb$ and $PbPb$ collisions at the LHC energy are presented. Here we
only list the parameter values used in the present calculation and show the
final results.

In our calculations for the nuclear density we take the standard Woods-Saxon
formula with the nuclear radius $R_{Pb}=6.62$ fm and the skin depth $d=0.546$
fm \cite{R}.

For the total inelastic $pp$ cross section at $\sqrt{s}=17.3$ GeV we take the
value obtained by the NA49 collaboration $\sigma_{\text{in}}=31.46$ mb
\cite{na49}. We assume the differential inelastic $pp$ cross section
$\sigma_{\text{in}}(s)$ to be in a Gaussian form with $\sigma_{\text{in}%
}(0)=0.92$ \cite{zero}.\footnote{We checked that different values of
$\sigma_{\text{in}}(0)$ hardly influence final results.}

The average number of wounded quarks and diquarks in a single $pp$ collision
$w_{p}^{(q+d)}=1.18$ (per one colliding proton)\footnote{At $\sqrt{s}=23$ GeV
$w_{p}^{(q+d)}\approx1.182$ and changes very slowly with energy \cite{bb-AA}%
.}. Finally we take $p_{q}=$ $w_{p}^{(q+d)}/3$ and $p_{d}=2p_{q}$ where
$p_{q}$ and $p_{d}$ are the probabilities for a quark and a diquark to
interact in a single $pp$ collision, respectively.\footnote{As discussed in
\cite{lhc} the specific relation between $p_{d}$ and $p_{q}$ is not important.}

\begin{figure}[h]
\begin{center}
\includegraphics[scale=1.2]{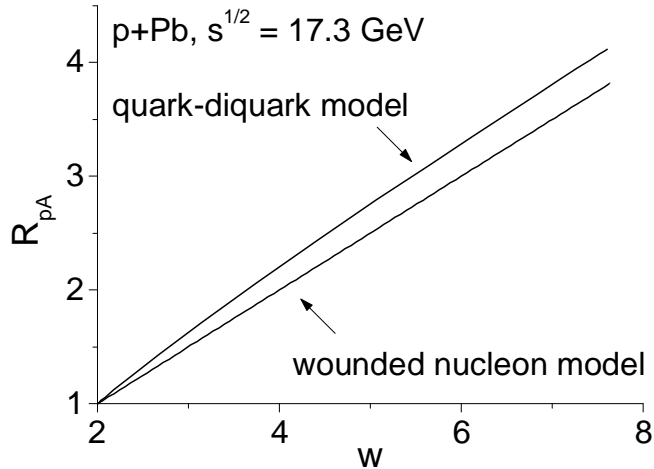}
\end{center}
\caption{Prediction of the wounded quark-diquark model for the midrapidity
ratio $R_{pA}$ compared with prediction of the wounded nucleon model.}%
\label{fig_ppb_ratio}%
\end{figure}

\textbf{3.} In Fig. \ref{fig_ppb_ratio} we present the predicted relation
between $R_{pA}\equiv N_{pA}(0)/N_{pp}(0)$ and the number of wounded nucleons
$w$ \cite{wnm}. $N_{pA}(0)$ and $N_{pp}(0)$ are the midrapidity particle
densities measured in $pPb$ and $pp$ collisions,
respectively.

\begin{figure}[h]
\begin{center}
\includegraphics[scale=1.2]{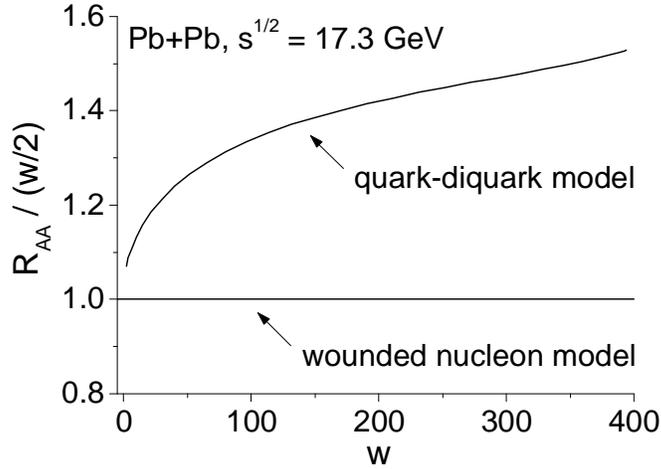}
\end{center}
\caption{Prediction of the wounded quark-diquark model for the ratio
$2R_{AA}/w$ compared with prediction of the wounded nucleon model.}%
\label{fig_pbpb_ratio}%
\end{figure}

In Fig. \ref{fig_pbpb_ratio} the wounded quark-diquark model prediction for
the ratio $R_{AA}/(w/2)$ vs the number of wounded nucleons $w$ is presented.
$R_{AA}\equiv N_{AA}(y)/N_{pp}(y)$ where $N_{AA}(y)$ is the particle density
measured in $PbPb$ collision. As explained in \cite{bb-AA,lhc} this ratio does
not depend on $y$, unless we are close to the fragmentation regions. It would
be very interesting to verify this strong consequence of the model when the
data are available.

\begin{figure}[h!]
\begin{center}
\includegraphics[scale=1.09]{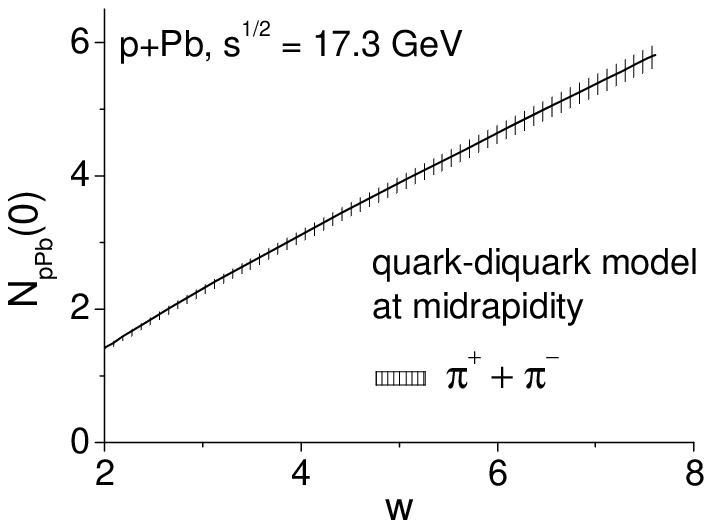}  \hskip0.3cm
\includegraphics[scale=1.09]{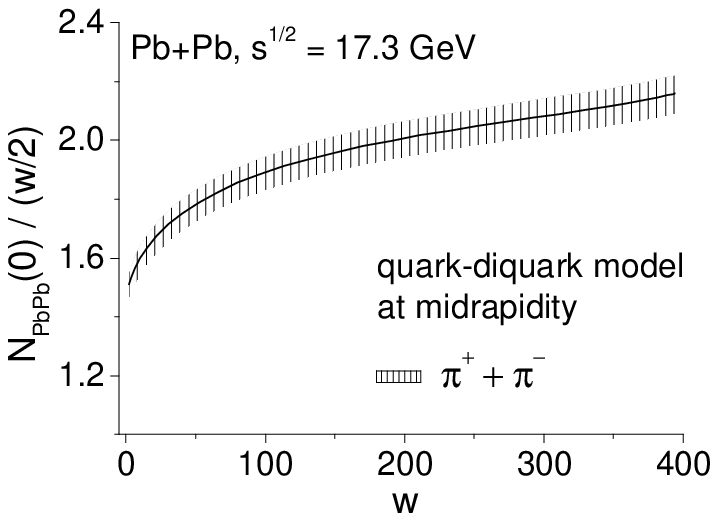}
\end{center}
\caption{Predictions of the wounded quark-diquark model for the charged pions
midrapidity densities in $pPb$ and $PbPb$ collisions. The error bars reflect
the inaccuracy in the $pp$ data.}%
\label{fig_end}%
\end{figure}

Multiplying $R_{pA}$ and $R_{AA}/(w/2)$ by the charged pions, $\pi^{+}+\pi
^{-}$,\footnote{In the present approach we cannot provide the separate
predictions for $\pi^{+}$ and $\pi^{-}$ multiplicities. At $\sqrt{s}=17.3$ GeV
the ratio $\pi^{+}/\pi^{-}$ is strongly influenced by the isospin effect
\cite{ar}.} midrapidity density in $pp$ collisions $N_{pp}(0)|_{\pi^{+}%
+\pi^{-}}=1.413$ (with the reasonable uncertainty of $3\%$) \cite{na49} we
obtain our final predictions for the charged pions midrapidity densities
presented in Fig. \ref{fig_end}. 

\bigskip

\textbf{Acknowledgements}

We would like to thank A. Rybicki for useful discussions on the NA49 data.
This investigation was supported in part by the Polish Ministry of Science and
Higher Education, grant No. N202 034 32/0918.

\end{document}